\documentclass[twoside,slac_one]{revtex4}
\usepackage{graphicx}
\usepackage{fancyhdr}
\usepackage{amsmath} 
\usepackage{bm}
\usepackage{amsxtra}
\usepackage{amssymb}
\usepackage{amsthm}
\usepackage{latexsym}
\usepackage{lscape}

\begin{document}

\title{Measurement of the Production Cross Section of Pairs of Isolated Photons at~$\sqrt{s}=7$~TeV with CMS}

\author{L. Millischer}
\affiliation{CEA Saclay, Gif-sur-Yvette, France}

\begin{abstract}
In this paper we present the measurement of the integrated and differential production cross sections of pairs of prompt isolated photons in proton-proton collisions at a centre-of-mass energy of 7~TeV with the Compact Muon Sollenoid~(CMS) detector at the Large Hadron Collider~(LHC). A next-to-leading order~(NLO) perturbative QCD~(pQCD) prediction was compared to the measurement, which was performed on a sample corresponding to an integrated luminosity of $36~\mbox{pb}^{-1}$. While the agreement on integrated cross sections is satisfactory, a discrepancy is observed in the region of the phase space populated by photons with small relative angle, where the theoretical prediction underestimates the measured cross section.
\end{abstract}

\maketitle

\thispagestyle{fancy}

\section{Introduction}
\label{sec_intro}

Studying the production cross section of pairs of energetic isolated photons at the LHC allows to test and validate pQCD calculations up to scales never probed before. Pairs of prompt photons, produced in the hard scattering of quarks and gluons, as opposed to photons produced in the decay of hadrons, also constitute the largest irreducible background to searches for a low mass Higgs boson decaying to two photons. 
\\

In this paper we present the measurement of the integrated and differential production cross sections of pairs of prompt isolated photons performed with the CMS detector~\cite{cms} at the LHC, at a centre-of-mass energy of 7~TeV using the data collected in 2010 corresponding to an integrated luminosity of $36~\mbox{pb}^{-1}$. The cross sections are measured for two isolated photons with transverse energy greater than 23 and 20~GeV respectively, separated by more than $R=0.45$ in the $(\eta,\varphi)$-plane and entering the pseudorapidity acceptance $|\eta|<1.44$ or $1.57<|\eta|<2.5$. The isolation criterion is defined at parton level, we require less than 5~GeV of transverse energy to be deposited in a cone of radius $R=0.4$ in the $(\eta,\varphi)$-plane. The differential cross sections are measured versus four variables, chosen either because of their sensitivity to the underlying QCD processes or their ability to describe the irreducible background in Higgs boson searches: the invariant mass $m_{\gamma\gamma}$, the transverse momentum of the photon pair $p_{T, \gamma\gamma}$, the azimuthal angle difference between the two photons $\Delta\varphi_{\gamma\gamma}$ and the cosine of the scattering angle in the Collins--Soper frame $\mbox{cos }\theta^*$~\cite{colsop}.
\\

After describing the event selection in section~\ref{sec_eventsel} we describe in section~\ref{sec_discr} a novel method to discriminate prompt an non-prompt photons and give the details of the cross-section measurement in section~\ref{sec_xsec}. The systematic uncertainties are summarised in section~\ref{sec_syst} and the results of a NLO prediction compared to the measured cross section in section~\ref{sec_results}. Pairs of photons are called \emph{diphotons}.

\section{Event Selection}
\label{sec_eventsel}

The CMS electromagnetic calorimeter (ECAL) consists of a cylindrical barrel part closed by two disc-shaped parts on either side, called \emph{endcaps}. It is made of about 76000 lead tungstate crystals, that are slightly off-pointing with respect to the interaction point. Photons produced in CMS deposit their energy in ECAL, photon candidates are reconstructed by building clusters of ECAL crystals~\cite{cluster}.
\\

Events with two photons having transverse energies greater than 23 and 20~GeV respectively, separated by more than $R=0.45$ and the clusters of which have $|\eta|<1.44$ (ECAL barrel) or $1.57<|\eta|<2.5$ (ECAL endcap) are selected. In order to reject non-prompt photons produced in jets, from the decay of energetic neutral mesons for instance, both photons are required to satisfy isolation criteria on the total track momentum and the total energy measured by the hadronic calorimeter~(HCAL) in a cone around the photon candidate as well as criteria on the cluster shape. Furthermore it is required that no \emph{impinging track}, defined as a track from the hard interaction with transverse momentum greater than 3~GeV, hits ECAL at less than $R=0.4$ of the photon. In order to reject isolated electrons misidentified as photons, events with a hit in the first layer of the tracking detector matched to the photon cluster are rejected; the remaining contamination of the sample with Drell--Yan events is estimated from simulated events and corrected.

An \emph{acceptance correction factor} is computed in each bin of the measured spectra, to account for the the finite detector resolution on the transverse energy and pseudorapidity, and the photon reconstruction and identification efficiencies. The efficiency of the isolation and cluster-shape criteria have been determined from data, using the \emph{tag and probe} method~\cite{tnp} on electrons from Z~boson decays. The efficiency of the selection on \emph{impinging tracks} was determined with \emph{random cones}, defined as a cone of $R=0.4$ around a direction at the same $\eta$ as the photon candidate and at a random $\varphi$ in a $\pi/2$ window around the axis perpendicular to the photon direction, since the probability for an \emph{impinging track} to hit a \emph{random cone} is the same as the probability for an \emph{impinging track} to hit the isolation surface of prompt photon. The acceptance correction factor is $76.2 \pm 3.3 \%$ in the region corresponding to the entire pseudorapidity range.

\section{Signal and Background Discrimination}
\label{sec_discr}

A sample of diphoton candidates selected with the requirements described in section~\ref{sec_eventsel} still contains a significant amount of background, that is diphoton candidates with one or both photons being non-prompt. In order to statistically determine the number of prompt isolated diphotons, we use the discrimination power of ECAL isolation variable $\mathcal{I}$, defined as the sum of the transverse energies of crystals with $E_T>300~\mbox{MeV}$ in an area around the photon ECAL impact position of outer radius 0.4, inner radius 3.5~crystal widths. Crystals belonging to the photon cluster or falling within a 5~crystal strip along $\varphi$ are removed from the sum. 
\\

Given that prompt isolated photons are not produced along electromagnetic particles, and given that the strip along $\varphi$ excludes the deposits of potential conversion products, they have much smaller ECAL isolation values than non-prompt photons, that are produced in jets along other electromagnetic particles and the deposits of which tend to leak into the isolation region by multiple conversion and bremsstrahlung processes. The distributions for prompt an non-prompt photons of the variable $\mathcal{I}$ are significantly different and can therefore be used in a maximum likelihood fit to extract the number of prompt diphotons event in any given sample of diphoton candidates, as shown in section~\ref{sec_xsec}. The extraction of the two distributions is entirely data-driven.
\\

\begin{figure}[ht]
\centering
\includegraphics[width=80mm]{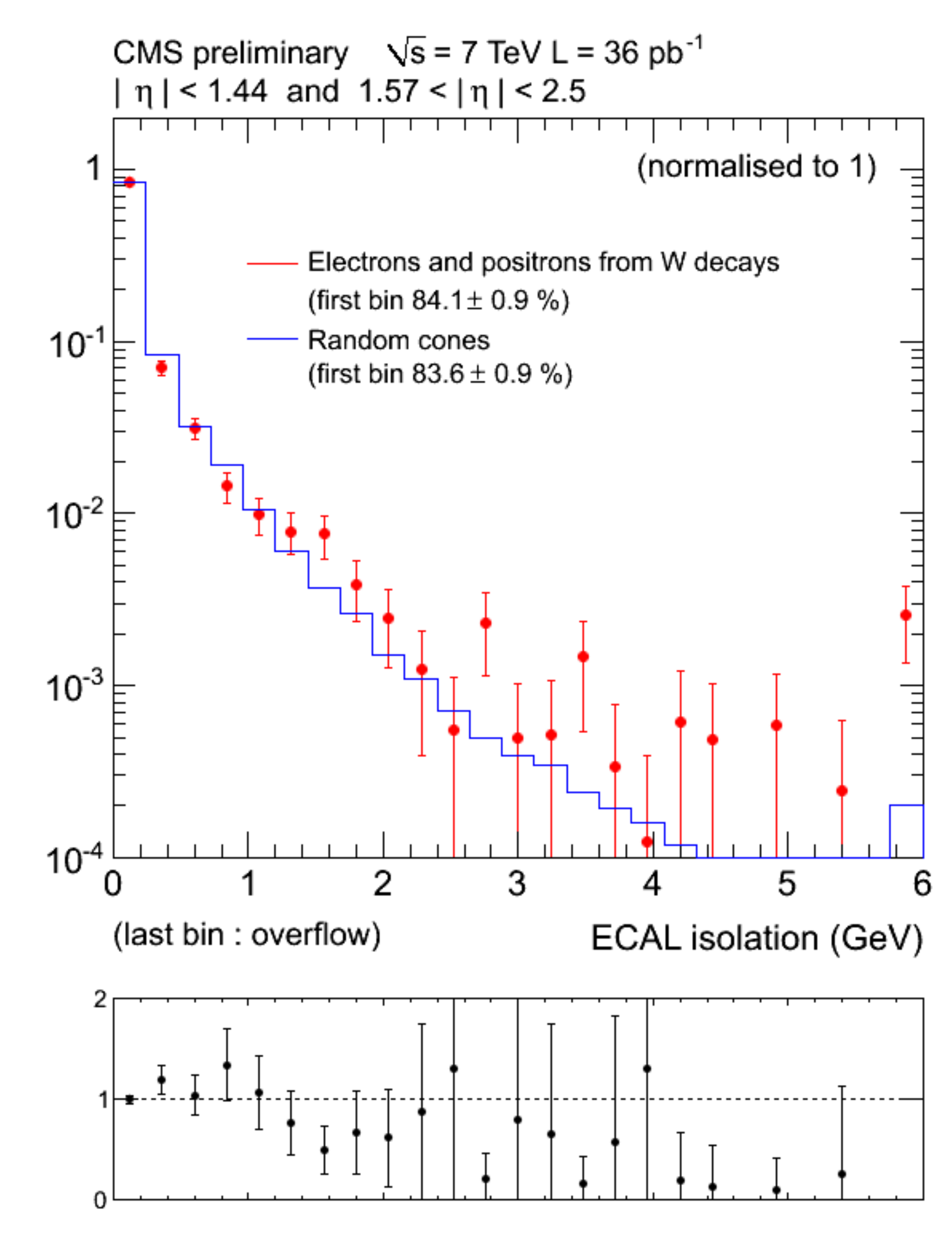}
\includegraphics[width=80mm]{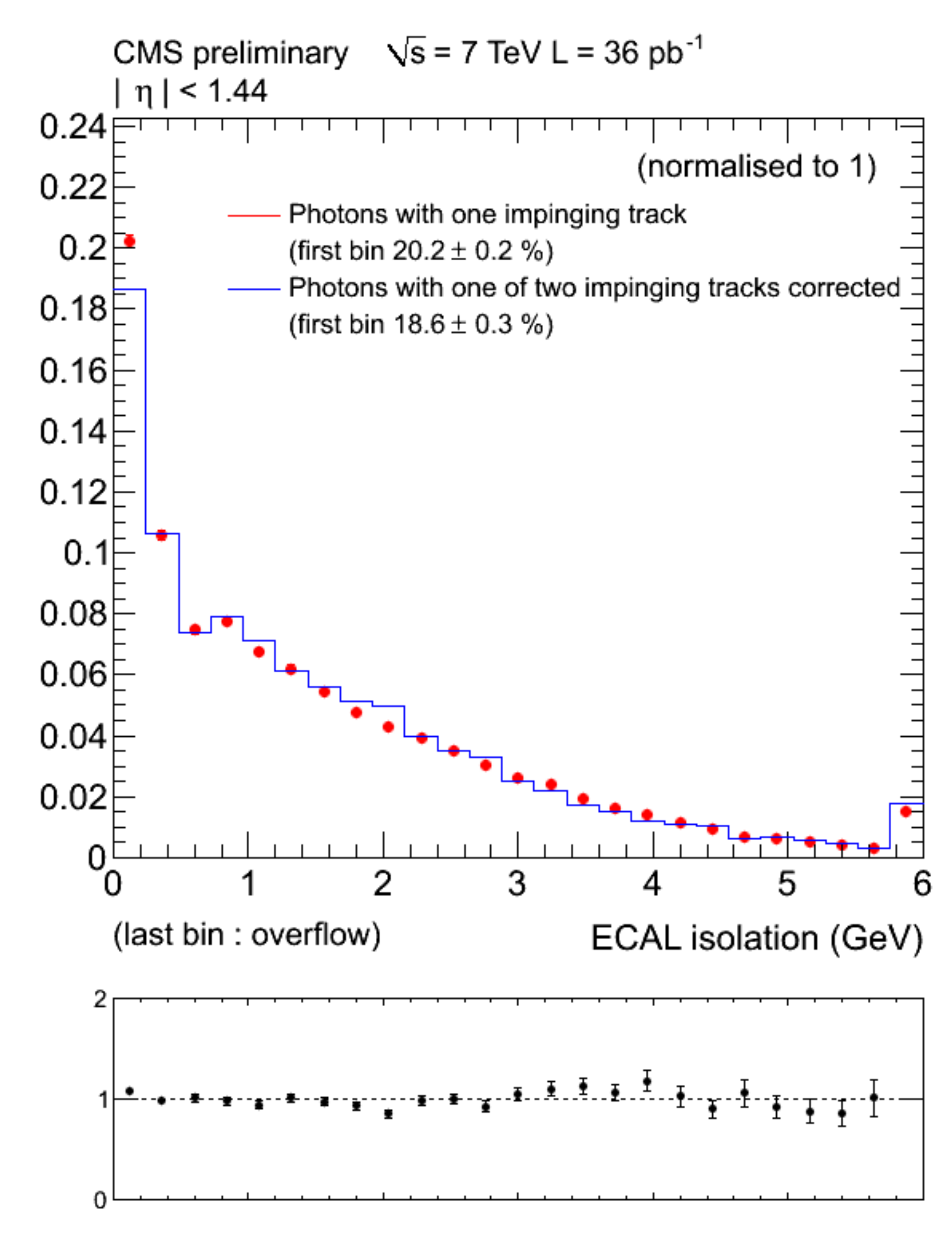}
\caption{Validation of the $\mathcal{I}$ shape extraction on data for prompt~(left) and non-prompt~(right) photons.} 
\label{fig_valid}
\end{figure}

The only contributions to the ECAL isolation of prompt photons come from pileup activity and the underlying events. These contributions being independent of $\varphi$, the distribution of the variable $\mathcal{I}$ of prompt photons can be extracted with the \emph{random cones} defined in section~\ref{sec_eventsel}. This \emph{random cone} technique is validated with simulated diphoton events, as well as with data electrons from Z and W-boson decays, having radiated a small fraction of their initial energy in bremsstrahlung from interactions with the tracker and therefore leaving deposits in ECAL similar to those of prompt isolated photons. For $W\rightarrow e\nu$ events, the \emph{sPlot} technique~\cite{splot} was exploited, using a variable based on the missing transverse energy as discriminator. The comparison of the $\mathcal{I}$ distributions of W-decay electrons and \emph{random cones} is shown on figure~\ref{fig_valid}. 
\\

The distribution of the variable $\mathcal{I}$ for non-prompt photons is extracted using a control sample containing less than 0.1~\% of prompt photons, obtained from requiring one and one only \emph{impinging track} to hit the isolation surface of the photon candidate. Subsequently the energy deposited by the charged particle that left the \emph{impinging track} in a cone of radius~0.05 around its ECAL impact point is removed from the variable $\mathcal{I}$ which is then scaled to the nominal isolation area. This \emph{impinging track} method is validated with simulated non-prompt photon events from QCD processes and with photon candidates in data, selected to be accompanied by two \emph{impinging tracks}. The comparison of the $\mathcal{I}$ distributions of photon candidates with one \emph{impinging track} and photon candidates with two \emph{impinging tracks}, the deposits of one of which have been removed, are shown on figure~\ref{fig_valid}.
\\

Thus, using the data-driven \emph{random cone} and \emph{impinging track} methods, it is possible to extract the distributions of the variable $\mathcal{I}$ for prompt and non-prompt photons, in the ECAL barrel and the endcaps separately. The discrepancies observed in the validation procedure (Fig.~\ref{fig_valid}) will be taken into account in the systematic uncertainties. 

\section{Cross-Section Measurement}
\label{sec_xsec}

In each bin of the measured spectra, the number of prompt isolated diphotons is extracted by means of an extended maximum likelihood~\cite{xlike} fit where the two-dimensional probability density functions~(pdf) are defined as the product of the one-dimensional distributions $f(\mathcal{I})$ extracted with the methods described in section~\ref{sec_discr}, for each of the three \emph{types} of diphotons: two prompt photons (the \emph{signal}), two non-prompt photons and one prompt and one non-prompt photon (the \emph{background}), . The likelihood function which is maximised is given in equation~\eqref{eqn_like} where $\mathcal{T}$ are the three event types, $N$ is the the size of the sample, $N_t$ the numbers of events estimated in the fit for each type $t$, $N_{tot}$ the sum over the three event types and $f_t(\mathcal{I}^1,\mathcal{I}^2)$ is the probability for the ECAL isolation variables of the two photons to have values $\mathcal{I}^1$ and $\mathcal{I}^2$ for the given event type $t$.
\\

\begin{equation}
\label{eqn_like}
  \mathcal{L} =
  \frac{e^{-N^{\text{tot}}}}{N!} \prod_{i=1}^{N}\ \sum_{t \in
  \mathcal{T}}{N_{t}f_{t}(\mathcal{I}^{1}_i,\mathcal{I}^{2}_i})~,
\end{equation}

\begin{figure}[ht]
\centering
\includegraphics[width=80mm]{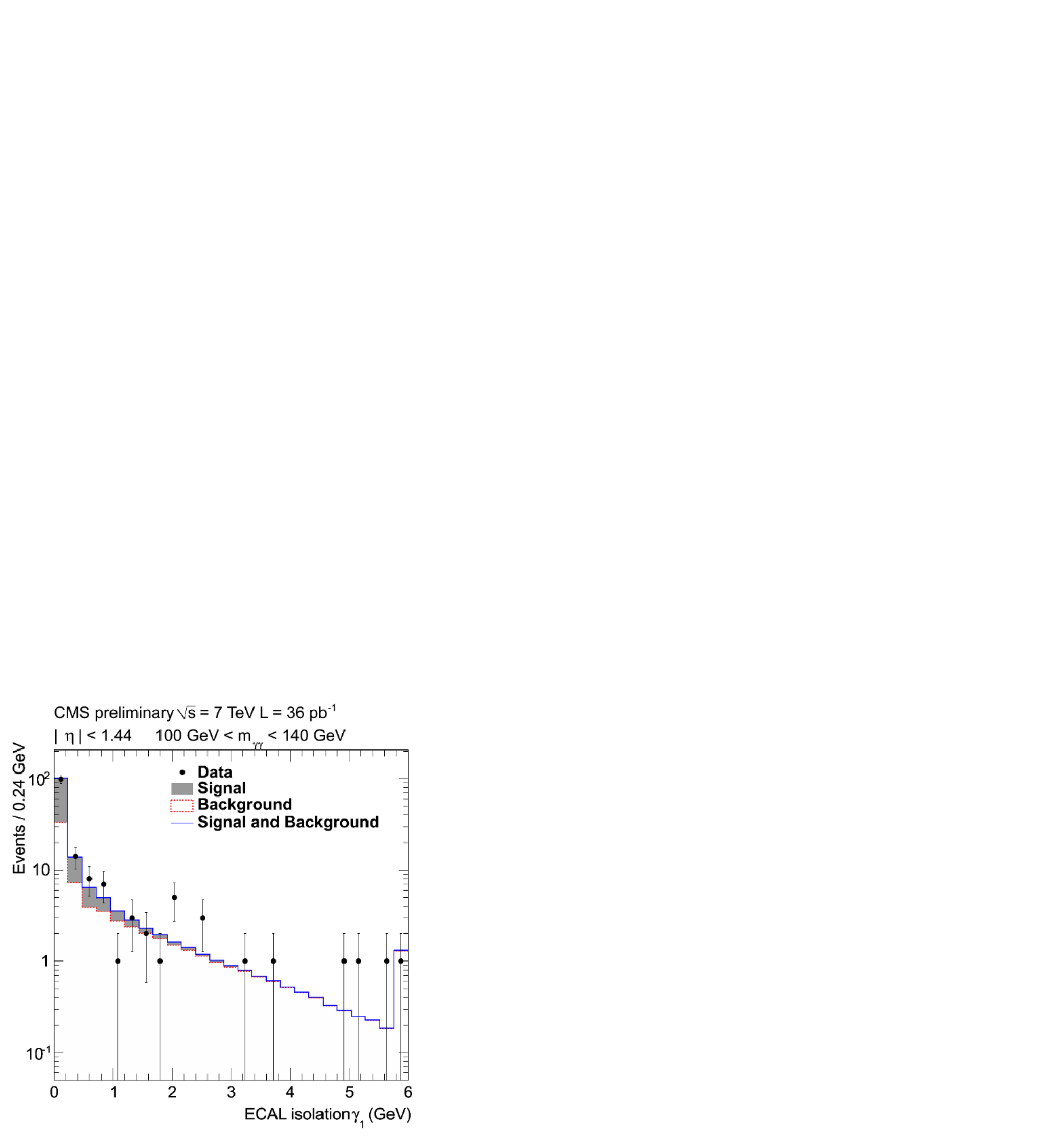}
\includegraphics[width=80mm]{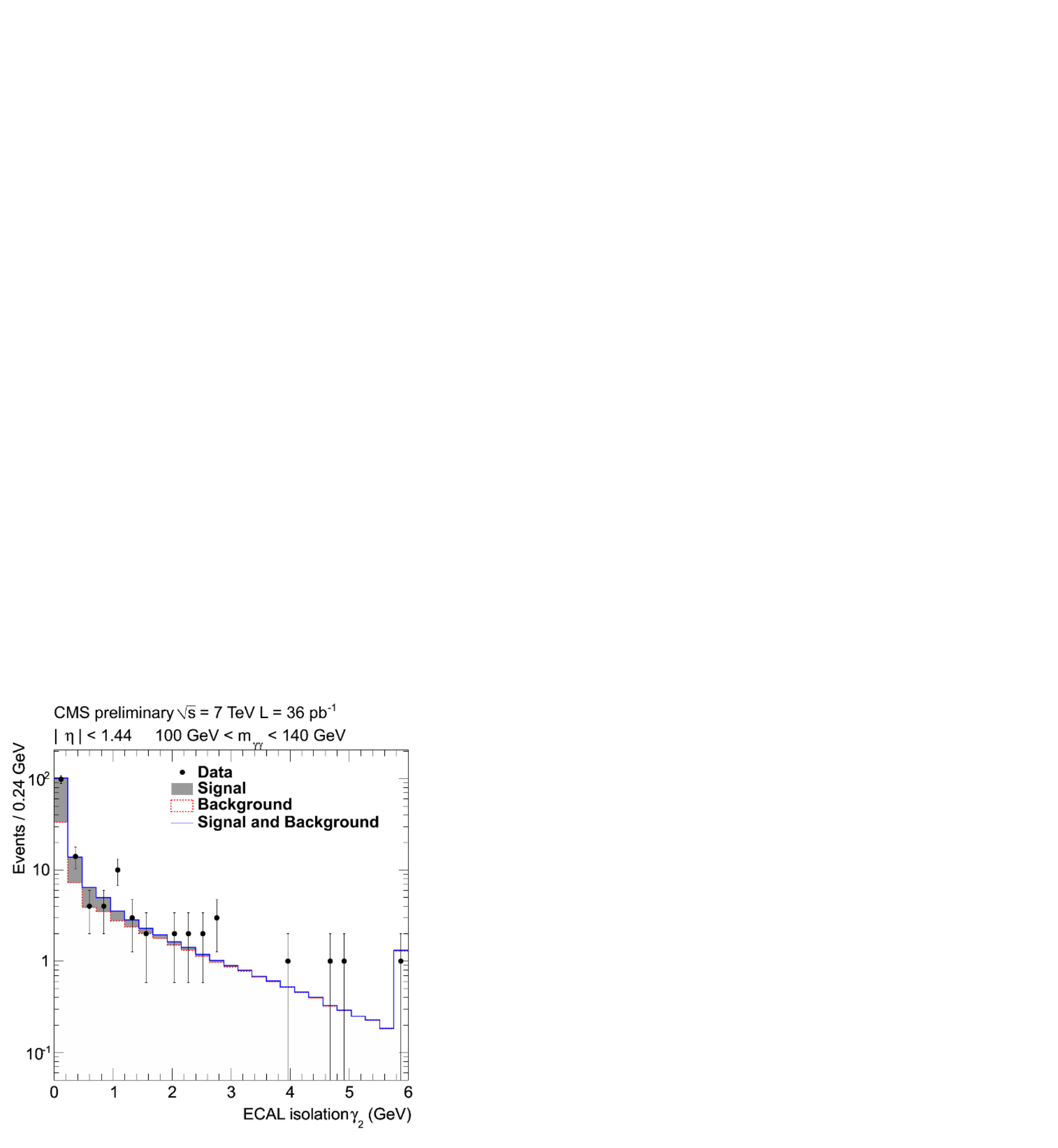}
\caption{Projections of the two-dimensional fit in the bin corresponding to $100~\mbox{GeV} < m_{\gamma\gamma} < 140~\mbox{GeV}$ for photons with $|\eta| < 1.44$. The continuous blue line represents the sum of the signal and the background contributions  the dashed red line represents the background contribution only. In this bin, 161 diphoton candidates were selected and the number of signal events has been determined to be $72 \pm 14$.} 
\label{fig_fit}
\end{figure}

Since the distributions of the variable $\mathcal{I}$ are extracted in the ECAL barrel and the endcap separately, the fit is performed for three different \emph{categories}: for events with both photons in the ECAL barrel, events with both photons in the ECAL endcap and events with one photon in each part of the ECAL. As the distributions $f(\mathcal{I})$ show a moderate dependence on $\eta$ and the pile-up and the distribution $f(\mathcal{I})$ of non-prompt photons also depends on the transverse energy $E_T$ of the candidate, the events in the samples used for the extractions of $f(\mathcal{I})$ are weighted so as to reproduce the distributions of $\eta$, $E_T$ and the number of pile-up vertices of the fitted diphoton sample. The fit performed in the bin corresponding to $100~\mbox{GeV} < m_{\gamma\gamma} < 140~\mbox{GeV}$ for both photons in the barrel is shown on figure~\ref{fig_fit}.
\\

The number of prompt isolated diphotons extracted with the maximum likelihood method is known to be a biased estimator of the actual number of signal events, for fitted samples with small numbers of events. This bias is estimated with Monte-Carlo pseudo-experiments, found to be smaller than half the statistical error in all bins and negligible in the bins with more than 100~events, and the result of the fit corrected for it.
The number of signal events estimated with the fit has been corrected for the reconstruction
and identification efficiencies via the acceptance correction factor described in section~\ref{sec_eventsel} and for the detector resolution by inverting the response matrix obtained from simulated events for $m_{\gamma\gamma}$, $p_{T,\gamma\gamma}$, $\Delta \varphi_{\gamma\gamma}$ and $\mbox{cos }\theta^*$. Given the excellent performance of ECAL, this matrix is nearly diagonal and no regularisation has been applied in the unfolding procedure.

\section{Systematic Uncertainties}
\label{sec_syst}

The imperfect knowledge of the signal and background pdfs used in the fit is the main source of systematic uncertainties, which are estimated using Monte Carlo pseudo-experiments with varied distributions. The amplitudes of the variation is taken to be the discrepancies between the distributions extracted with the methods described in section~\ref{sec_discr} and the cross-check distributions, examples of which are shown on figure~\ref{fig_valid}. They are of the order of $\pm 0.01$ for the signal and range from $\pm 0.03$ to $\pm 0.05$ for the background. 
\\

The ECAL energy scale, known to 0.6~\% in the barrel and 1.5~\% in the endcaps~\citep{escale}, affecting the definition of the acceptance via the $E_T$ thresholds defined in section~\ref{sec_eventsel} and induced bin-to-bin migrations in the differential spectra affected by the energy measurement, that is $m_{\gamma\gamma}$ and $p_{T, \gamma\gamma}$. The former uncertainty is particularly important in regions populated by photons close to the $E_T$ thresholds of 20 and 23~GeV.
\\

The uncertainties associated with the computation of the acceptance correction factor, defined in section~\ref{sec_eventsel}, are a combination of the statistical uncertainty on the samples used and the systematic uncertainties of the methods employed, taken to be the differences obtained when applying the method to samples of data and simulated events. The uncertainty on the integrated cross section is 3.3~\% for diphotons entering the entire pseudorapidity acceptance.
\\

\begin{figure}[ht]
\centering
\includegraphics[width=75mm]{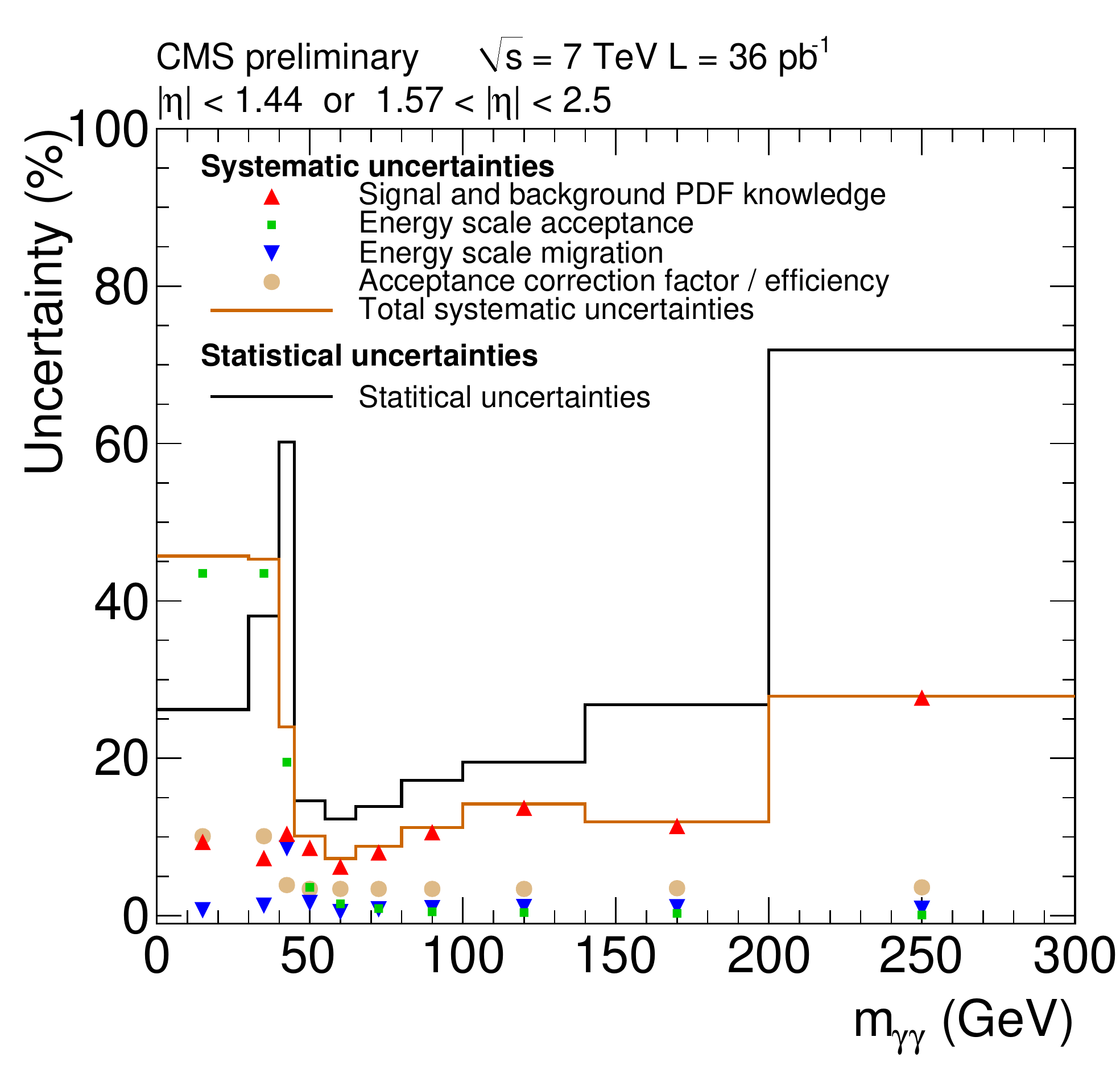}
\includegraphics[width=75mm]{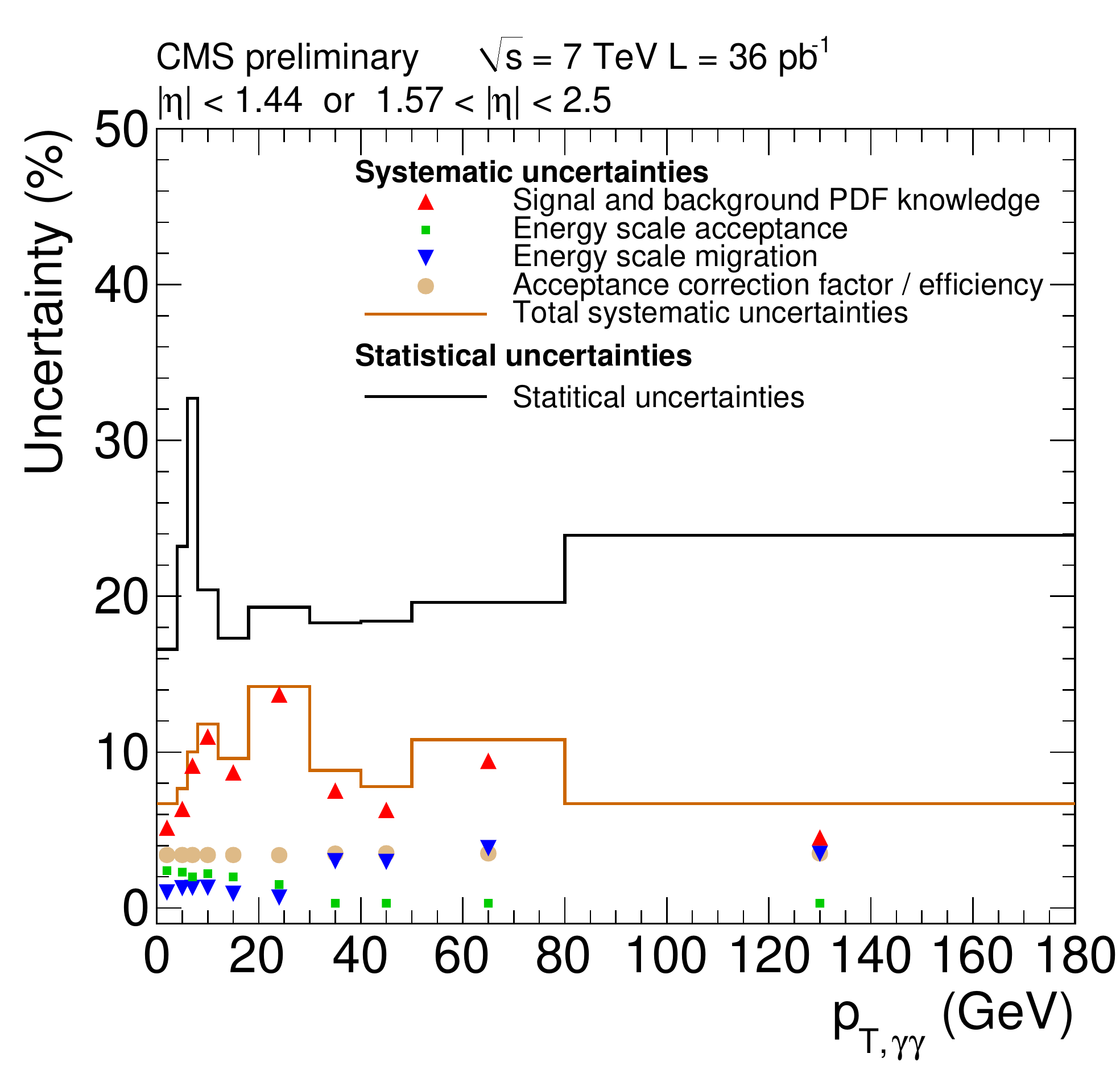}
\\
\includegraphics[width=75mm]{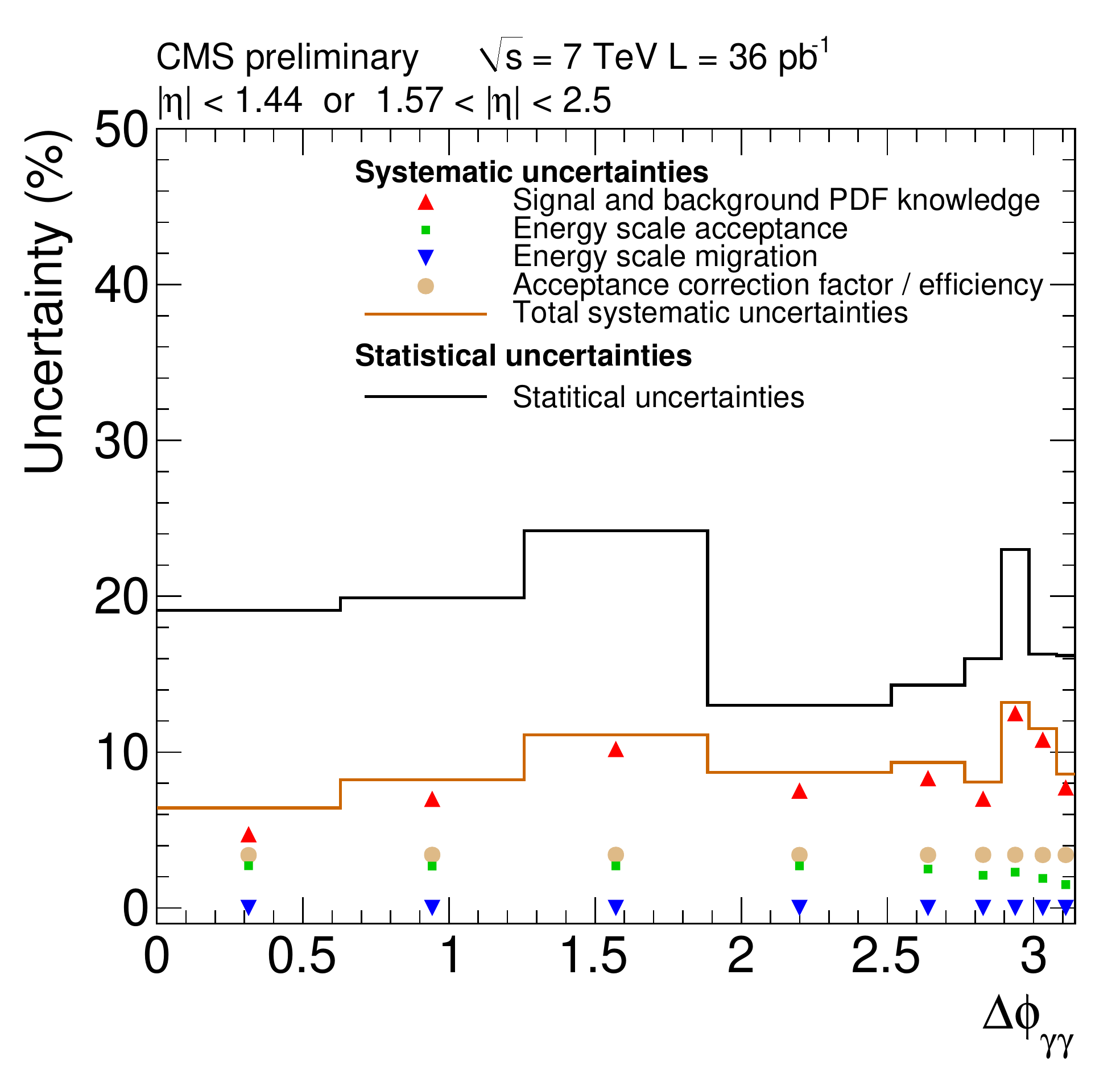}
\includegraphics[width=75mm]{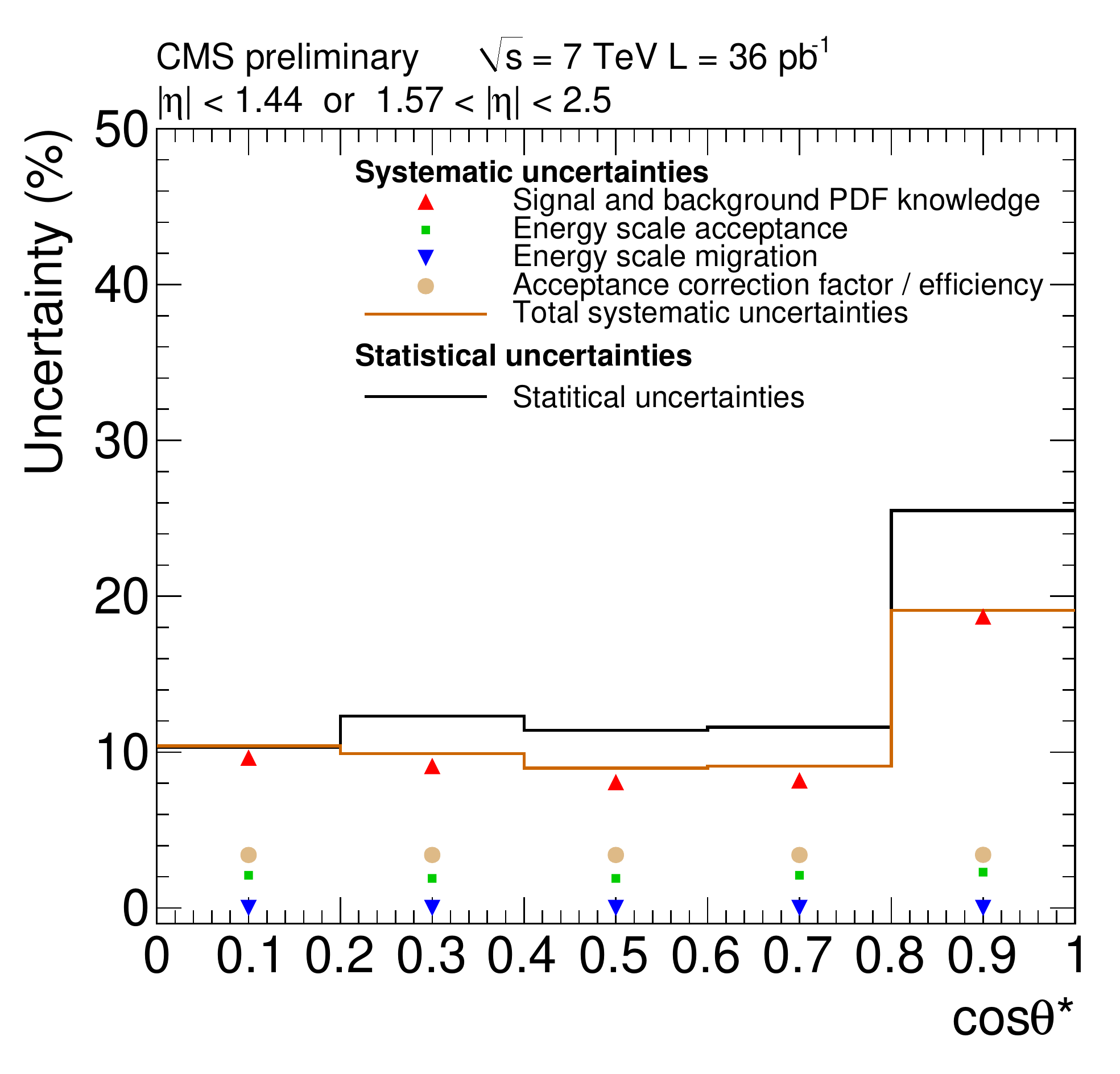}
\caption{Summary of the statistical and systematic uncertainties in the four measured spectra.} 
\label{fig_uncert}
\end{figure}

On top of these uncertainties, an uncertainty of 4~\% is assigned to the knowledge of the integrated luminosity of the dataset~\cite{lumi}. The systematic uncertainties on the four spectra are summarised on figure~\ref{fig_uncert}, showing that with the exception of the region corresponding to $m_{\gamma\gamma}<40~\mbox{GeV}$, the measurement is dominated by statistical errors. The systematic uncertainties are dominated by the uncertainty on the signal and background pdfs, except for the same mass region, populated by photons close to the $E_T$ threshold, leading to significant uncertainties on the acceptance definition via the knowledge of the ECAL energy scale.
\\

\section{Results}
\label{sec_results}

A fixed order theoretical prediction is compared with the measured spectra. It includes all contributing processes at NLO: the quark annihilation process and the single and double fragmentation processes at order $\alpha_s\alpha^2$ calculated with the {\sc diphox} parton generator code~\cite{diphox} and the gluon fusion process, enhanced at the LHC because of high gluon densities, up to and including the order $\alpha_s^3\alpha^2$ computed with the {\sc gamma2mc} generator code~\cite{g2mc}.
\\

As particles resulting from underlying event activity and hadronisation are not included in partonic event generators such as {\sc diphox} and {\sc gamma2mc}, the fraction of diphotons not selected due to underlying hadronic activity falling inside the isolation cone has been estimated using the {\sc pythia} 6.4~\cite{pythia} event generator. This results in the parton-level cross section being reduced by $4.7 \pm 0.3~\%$. Theoretical uncertainties on the knowledge of the parton distribution functions~(PDF) and $\alpha_s$ were determined following the PDF4LHC recommendations~\cite{pdf4lhc} and the factorisation, renormalisation and fragmentation scales were varied by factors 1/2 and 2, keeping the ratio between two scales less than 2. The predicted integrated cross section is:
\\

\begin{equation}
  \begin{array}{rcl}
    \sigma(pp\rightarrow\gamma\gamma)|_{|\eta| < 2.50}^{\text{pred.}} & = & 52.7 \ \ \ \mbox{ }^{ +5.8}_{ -4.2} \ \ \text{(scales)} \ \ \ \pm 2.0 \ \ \text{(PDF)} \ \ \ \text{pb}.\\
  \end{array}
\end{equation}

It is found to be compatible with the measured integrated cross section within the statistical and systematic uncertainties:

\begin{equation}
  \begin{array}{rcl}
    \sigma(pp\rightarrow\gamma\gamma)|_{|\eta|< 2.50}^{\text{meas.}} &=& 62.4 \ \ \pm 3.6
    \ (\text{stat}) \ \ \mbox{ }^{ +5.3}_{ -5.8} \  (\text{syst}) \ \ \pm 2.5 \  (\text{lumi}) \ \ \ \text{pb}.\\
  \end{array}
\end{equation}

\begin{figure}[ht]
\centering
\includegraphics[width=80mm]{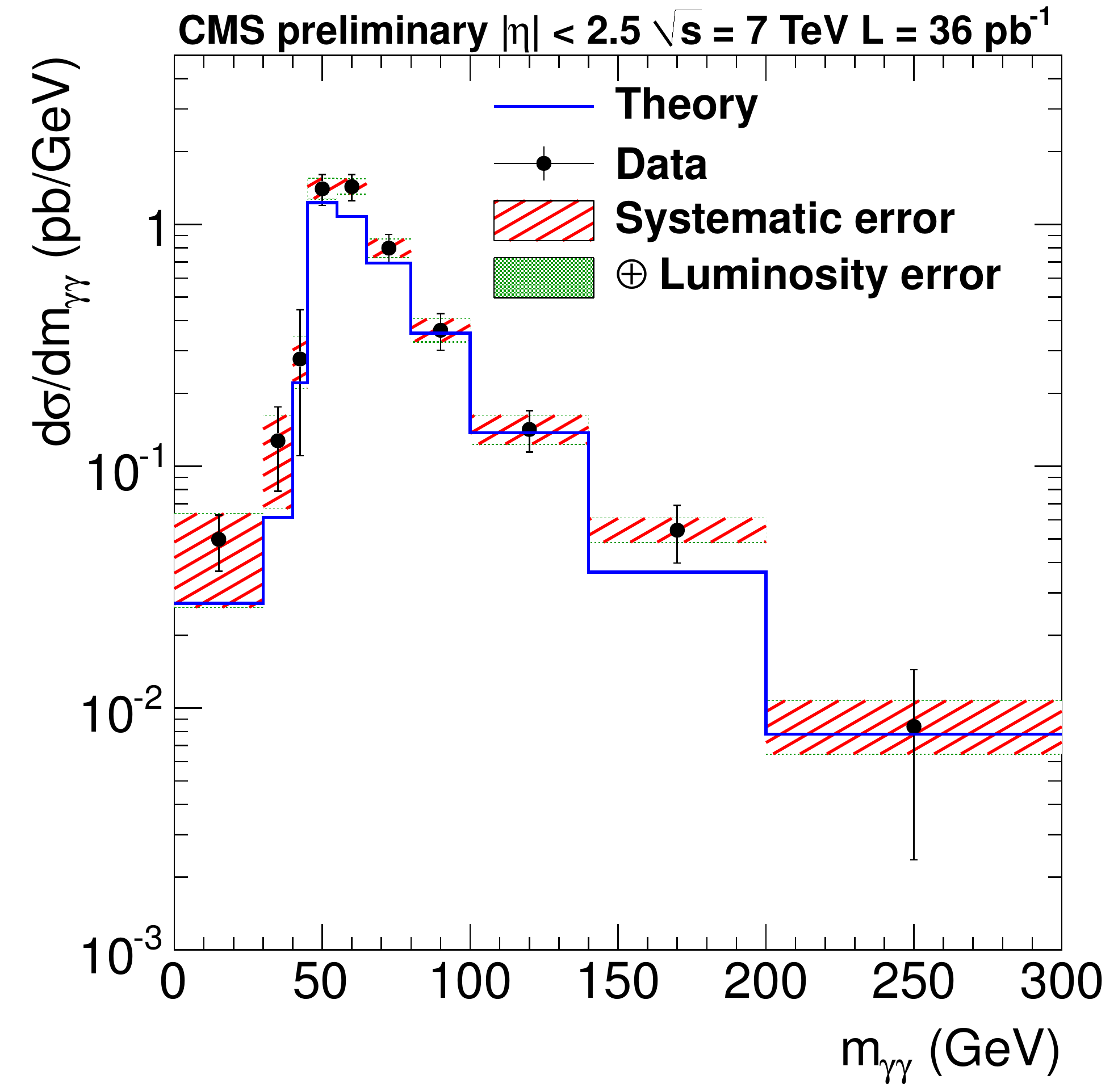}
\includegraphics[width=80mm]{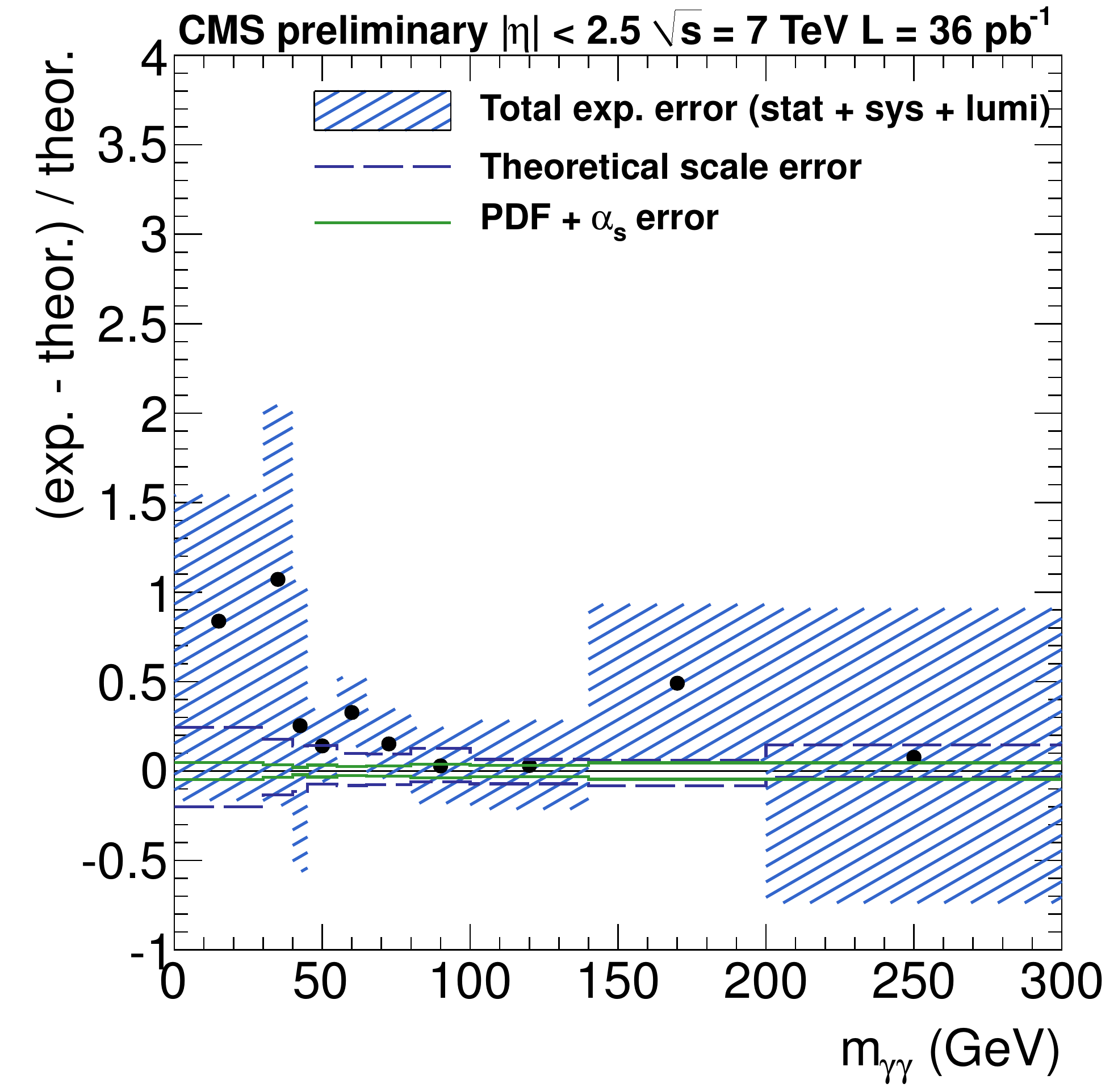}
\caption{Measured cross section of diphoton production as a function of the invariant mass
of the photon pair~(left) and bin-by-bin comparison with the theory~(right) for photons
within the pseudorapidity region $|\eta| < 1.44$ or $1.57 < |\eta| < 2.5$. The total systematic uncertainties are represented by the shaded area, the different contributions are added in quadrature sequentially.} 
\label{fig_result_m}
\end{figure}

\begin{figure}[ht]
\centering
\includegraphics[width=80mm]{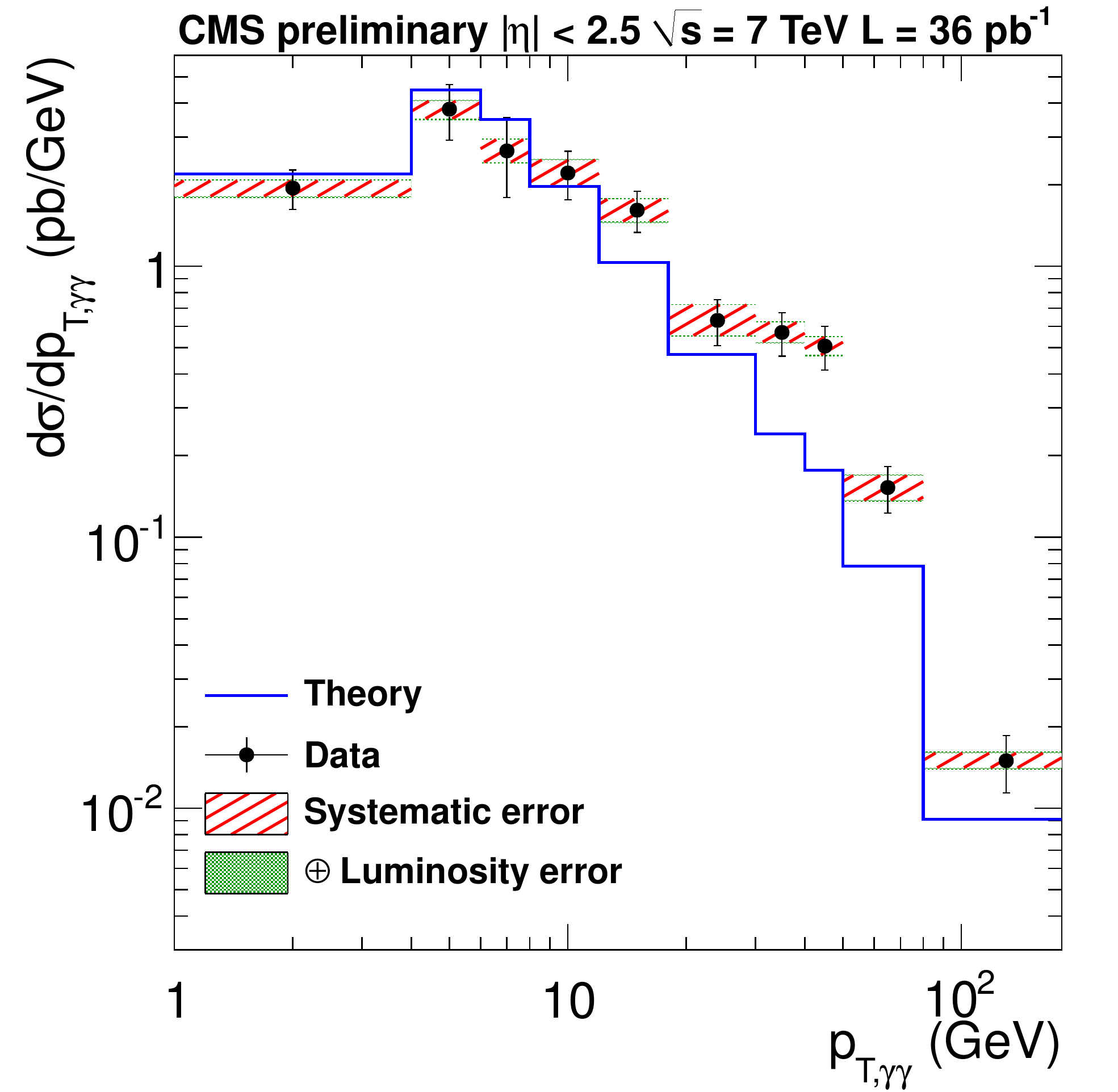}
\includegraphics[width=80mm]{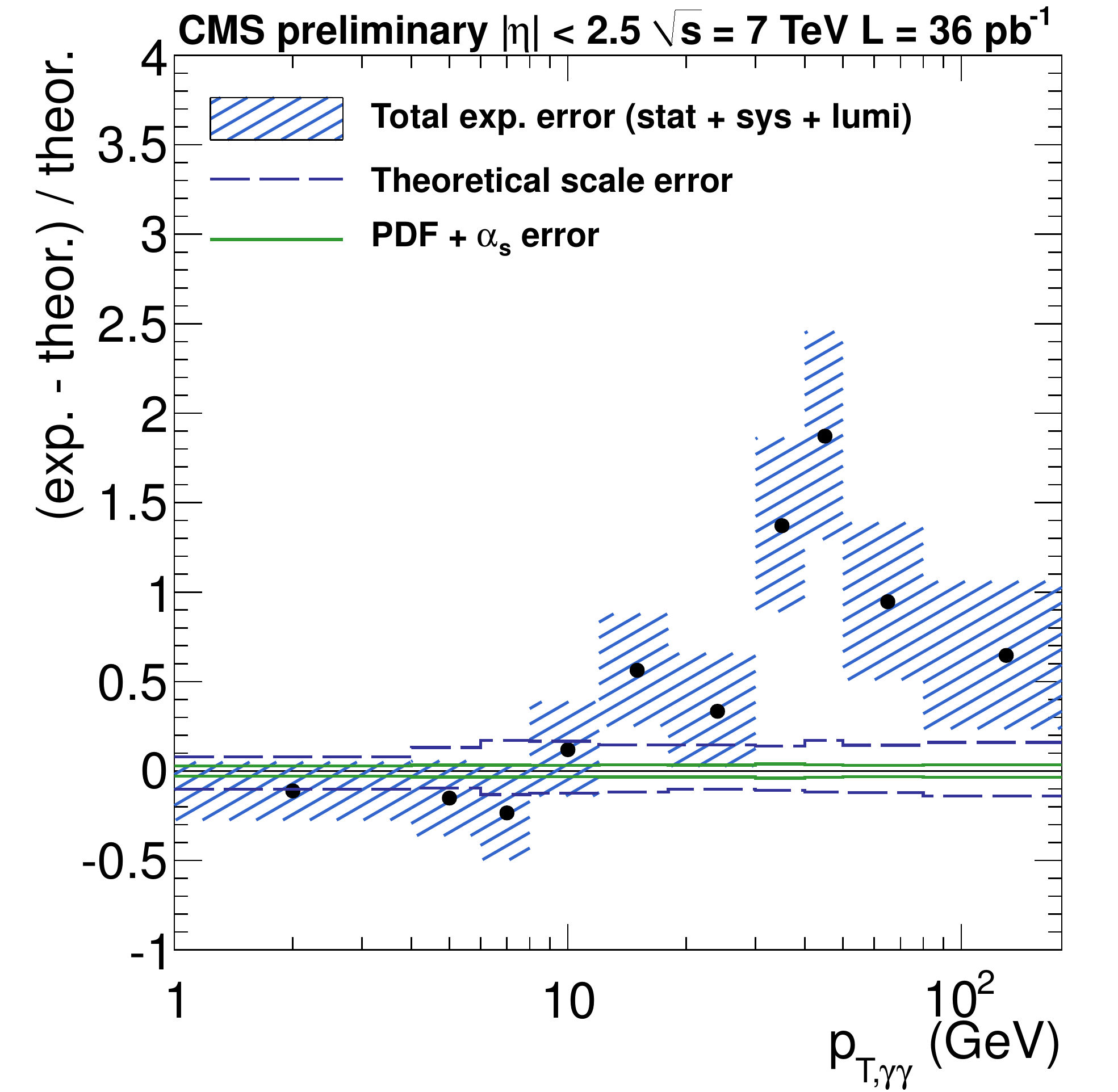}
\caption{Measured cross section of diphoton production as a function of the transverse momentum
of the photon pair~(left) and bin-by-bin comparison with the theory~(right) for photons within the pseudorapidity region $|\eta| < 1.44$ or $1.57 < |\eta| < 2.5$. The total systematic uncertainties are represented by the shaded area, the different contributions are added in quadrature sequentially.} 
\label{fig_result_pt}
\end{figure}

\begin{figure}[ht]
\centering
\includegraphics[width=80mm]{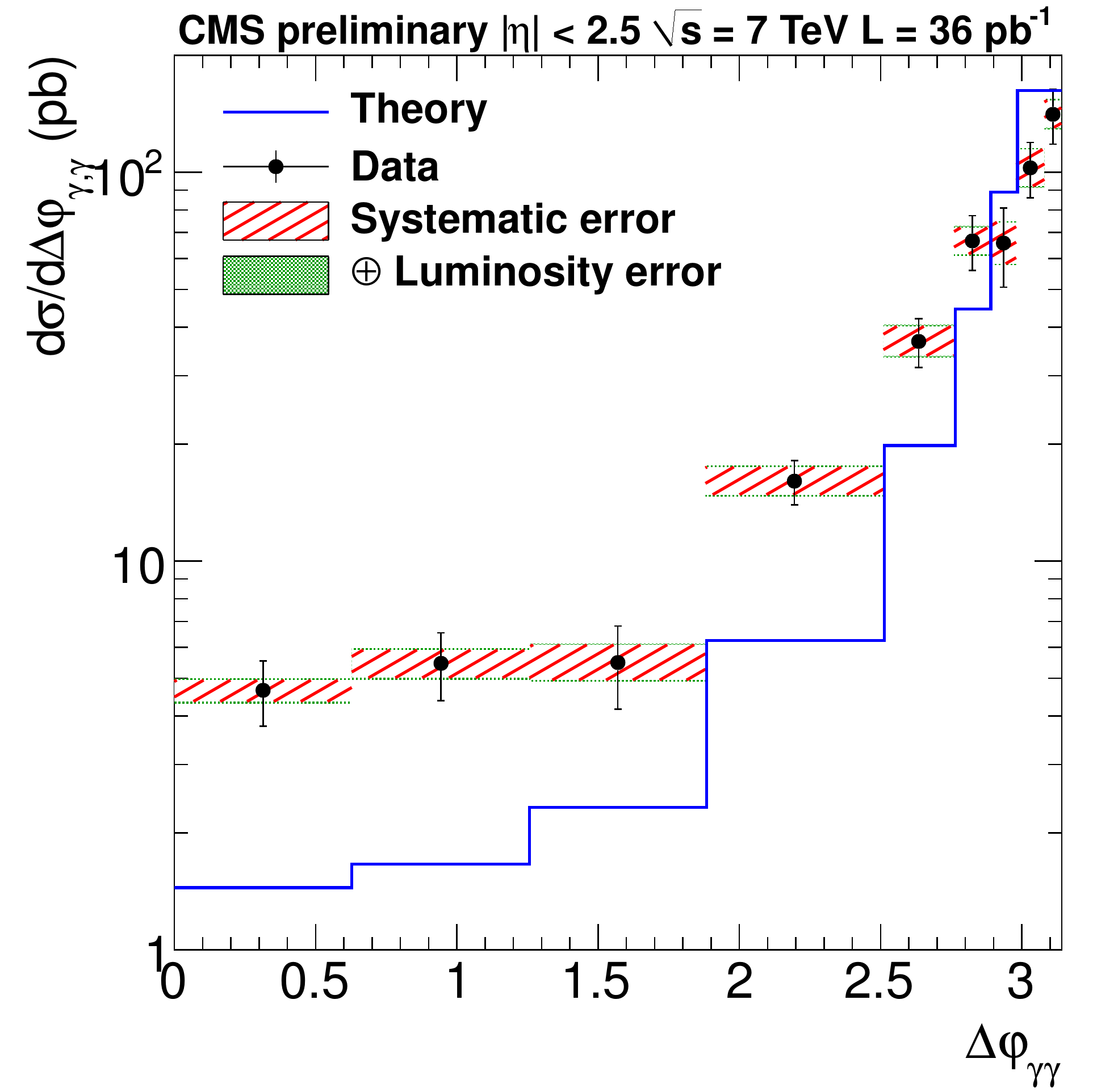}
\includegraphics[width=80mm]{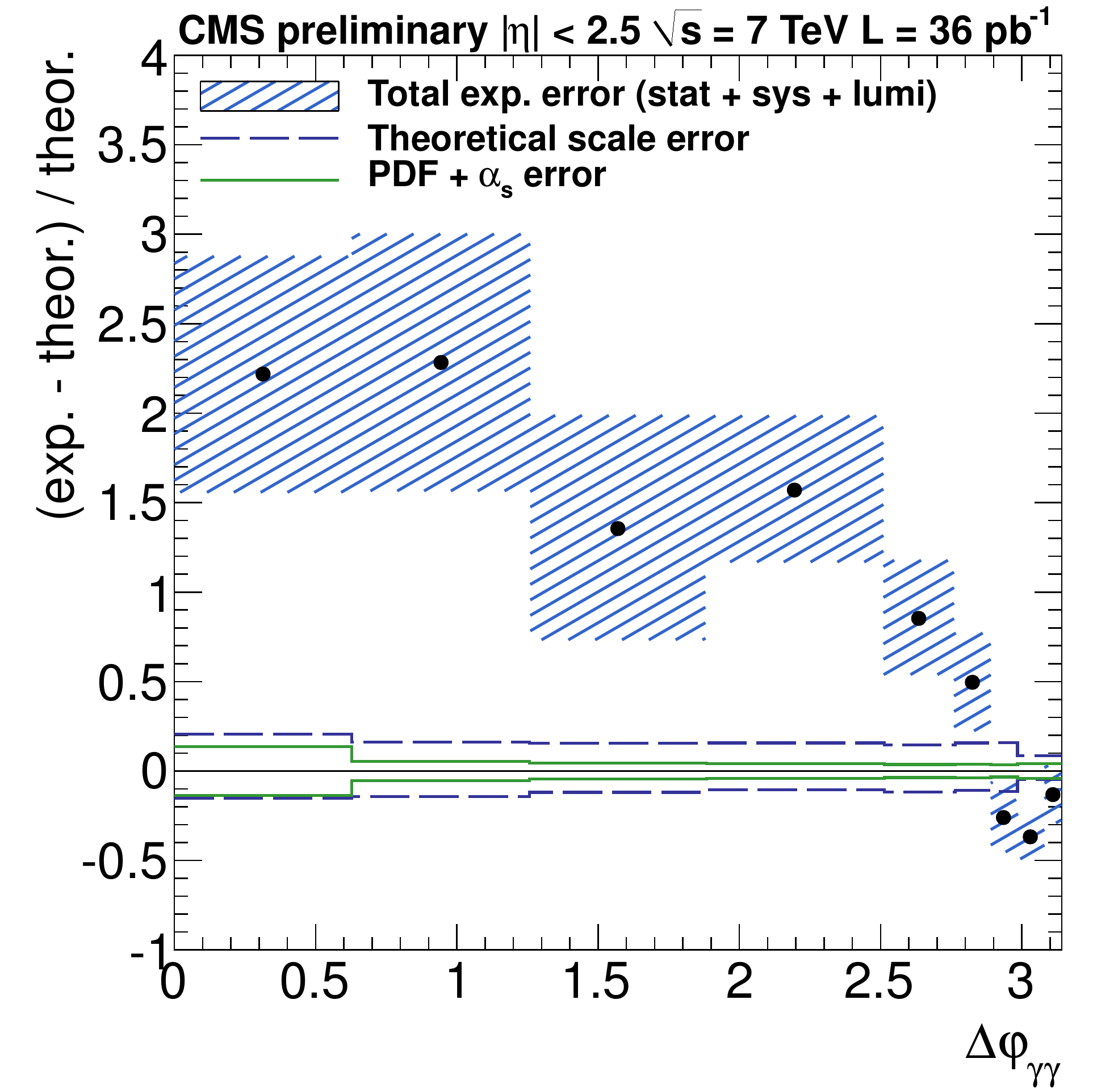}
\caption{Measured cross section of diphoton production as a function of the azimuthal angle between the two photons~(left) and bin-by-bin comparison with the theory~(right) for photons within the pseudorapidity region $|\eta| < 1.44$ or $1.57 < |\eta| < 2.5$. The total systematic uncertainties are represented by the shaded area, the different contributions are added in quadrature sequentially.} 
\label{fig_result_dphi}
\end{figure}

\begin{figure}[ht]
\centering
\includegraphics[width=80mm]{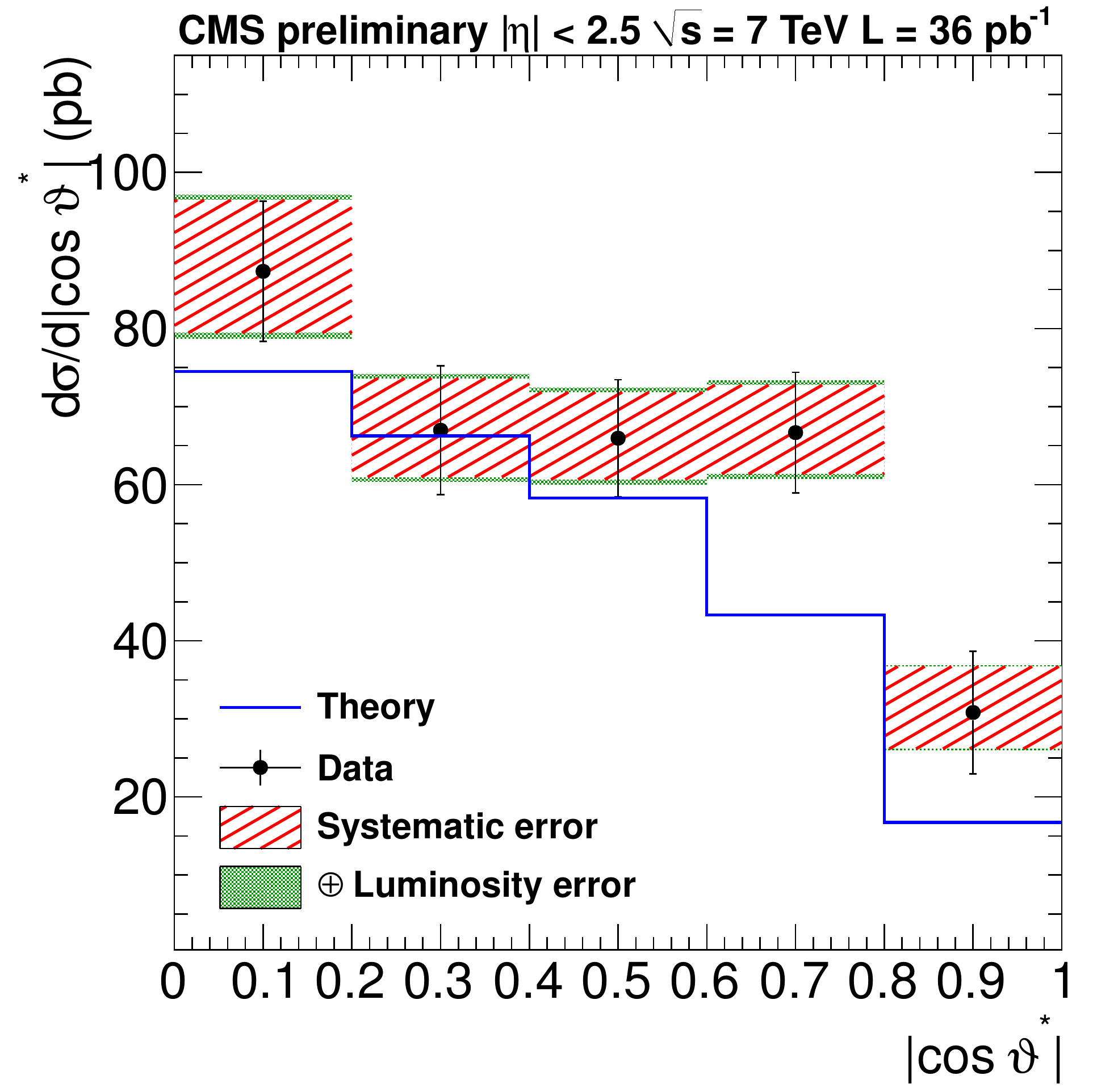}
\includegraphics[width=80mm]{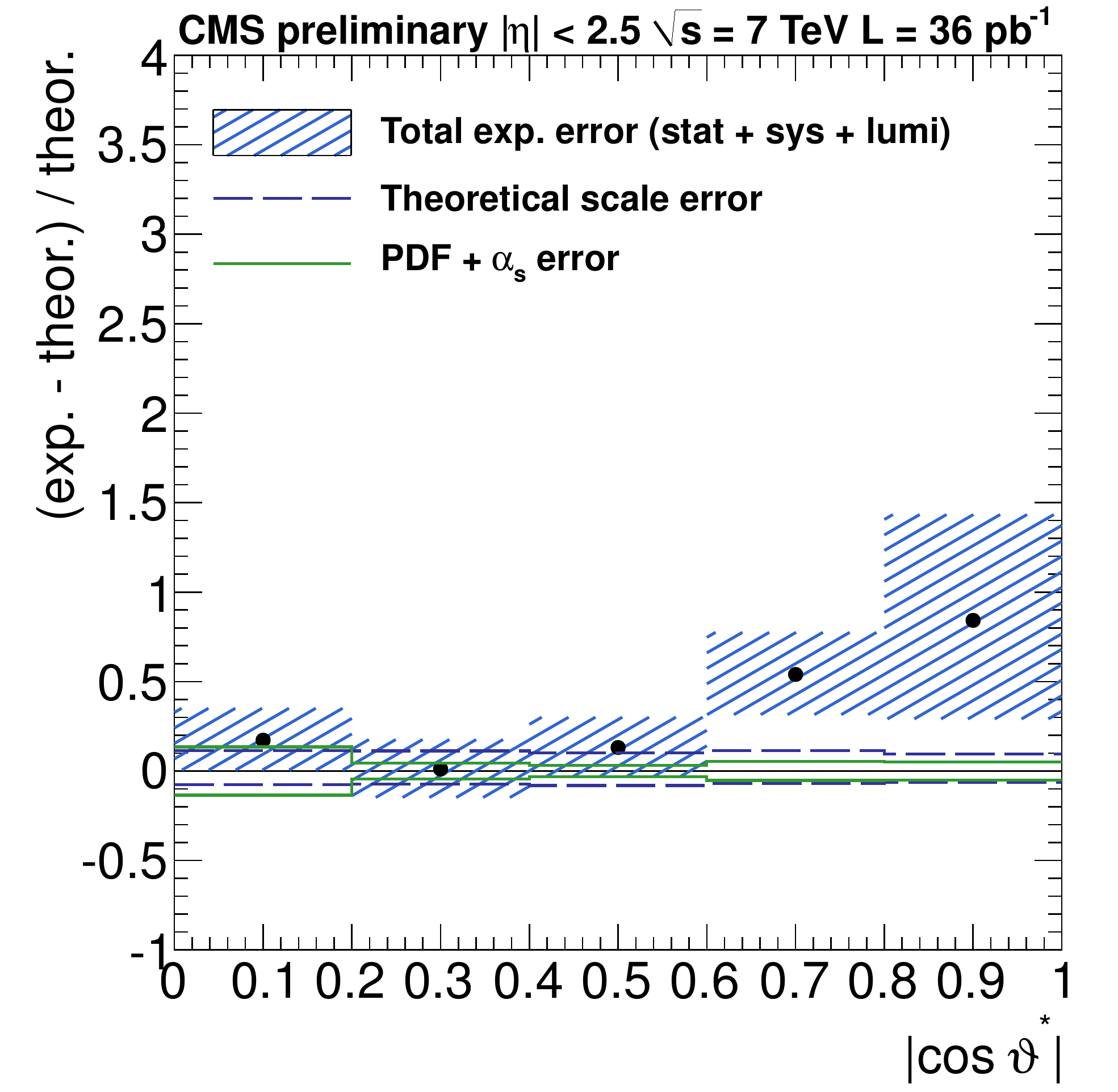}
\caption{Measured cross section of diphoton production as a function of the absolute value
of the cosine of the polar angle between the two photons in the Collins-Soper reference frame
and the photon pair~(left) and bin-by-bin comparison with the theory~(right) for photons within the pseudorapidity region $|\eta| < 1.44$ or $1.57 < |\eta| < 2.5$. The total systematic uncertainties are represented by the shaded area, the different contributions are added in quadrature sequentially.} 
\label{fig_result_cth}
\end{figure}

The predicted differential cross-section spectra for the variables $m_{\gamma\gamma}$, $p_{T, \gamma\gamma}$, $\Delta\varphi_{\gamma\gamma}$ and $\mbox{cos }\theta^*$ are compared to the measured spectra on figures~\ref{fig_result_m}, \ref{fig_result_pt}, \ref{fig_result_dphi} and~\ref{fig_result_cth} respectively.
\\

Figure~\ref{fig_result_dphi} shows that the theoretical cross section is underestimated in the region $\Delta\varphi_{\gamma\gamma}<2.8$. Indeed, in the leading order~(LO) description of the gluon fusion and quark annihilation processes, the two photons are back-to-back because of momentum conservation. Therefore the LO term does not contribute to this phase space region, which is effectively covered in the NLO calculation by one order only. This affects in particular the fragmentation contributions~\cite{shoulder}. Diphotons with small relative angle between the photons populate the regions $m_{\gamma\gamma}<40~\mbox{GeV}$ and $p_{T, \gamma\gamma}>40~\mbox{GeV}$ (because of the requirement of $E_T>20,23~\mbox{GeV}$) both of which are also underestimated by the theoretical prediction. 

Analogous disagreements were observed in the diphoton production in proton-antiproton collisions by the CDF~\cite{cdf} and D0~\cite{d0} collaborations as well as in proton-proton collisions by the ATLAS~\cite{atlas} collaboration.

\section{Conclusion}
\label{sec_conc}

The integrated and differential production cross sections of isolated photon pairs have been  measured in proton-proton collisions at a centre-of-mass energy of 7 TeV. The differential cross sections were measured as functions of the diphoton invariant mass and transverse momentum, the azimuthal angle difference between the two photons, and the cosine of the scattering angle in the Collins-Soper reference frame. The non-prompt photon background contamination from hadron decay products has been estimated with a statistical method based on an electromagnetic energy isolation variable, the signal and background distributions of which have been entirely extracted from data resulting in systematic uncertainties of approximately 10~\% on the measured cross section. The measurements have been compared to a theoretical prediction performed at NLO order accuracy using the state-of-the-art fixed order computations, showing that in regions of phase space mostly populated by photons emitted at small relative angle the prediction underestimates the measured cross section.


\bigskip
\bibliographystyle{ieeetr}
\bibliography{DPF_CMS_diphotons}



\end{document}